\documentclass[a4paper,fleqn]{cas-sc}

\usepackage[authoryear,longnamesfirst]{natbib}

\def\tsc#1{\csdef{#1}{\textsc{\lowercase{#1}}\xspace}}
\tsc{WGM}
\tsc{QE}
\tsc{EP}
\tsc{PMS}
\tsc{BEC}
\tsc{DE}

\begin{document}
\let\WriteBookmarks\relax
\def\floatpagepagefraction{1}
\def\textpagefraction{.001}
\shorttitle{Formation of Pareto distribution}
\shortauthors{SV Filatov et~al.}

\title [mode = title]{The formation of Pareto distribution in tracer systems on the water surface} 

\author[1,2]{S.V. Filatov}
\cormark[1]
\fnmark[1]
\ead{fillsv@issp.ac.ru}

\address[1]{142432, Institute of Solid State Physics, RAS}
\address[2]{142432, Landau Institute for Theoretical Physics, RAS}

\author[1,2]{A.V. Poplevin}

\author[1,2]{A.A. Levchenko}

\credit{Data curation, Writing - Original draft preparation}

\author[1,2]{E.V. Lebedeva}

\author[1,2]{L.P. Mezhov-Deglin}

\cortext[cor1]{Corresponding author}
\cortext[cor2]{Principal corresponding author}

\begin{abstract}
We experimentally studied a clusterization process in a system of polyamide tracers that are used for visualizing the flow of liquids on their surface. 
It was shown that in a surface structure system appearing on the water surface a Pareto distribution is formed for normalized cluster density and it is well described by a $C/x^{-n}$ power-law function. One can suggest that the growth of a surface structure is defined by the action of surface tension forces, and the number of surface structures decreases exponentially, while maintaining their total surface area. We show experimentally the significant role of background liquid flows on the surface in the clusterization process.
\end{abstract}

\begin{graphicalabstract}
\end{graphicalabstract}

\begin{highlights}
\item The interaction of neutral tracer particles on the water surface is provided by capillary forces.
\item The growth of large-scale structures is observed experimentally; a Pareto distribution is formed in the cluster system - the power-law distribution of the particle density over their areas.
\item The maximum cluster size is limited by random fluid flows.
\end{highlights}


\maketitle

\section{Introduction}

The experimental procedure of studying vortex currents created by waves on the surface of liquids includes visualization using a decorative powder with the density that is close to the density of the liquid \cite{VonKameke2011, Filatov2015, Francois2013, Mezhov-Deglin2019}. Normally a polyamide-12 powder with a mean size of $\sim$ 30 $\mu$m \citep{VonKameke2011, Filatov2015, Francois2013} or hollow glass spheres of about 50 $\mu$m in diameter [\cite{Mezhov-Deglin2019}] are used. The liquid surface is photographed from above by a camera with a certain shooting frequency. Compairing the sequence of photos allows computing the liquid's surface velocity by means of the PIVlab software package [\cite{Thielicke2014}]. It requires the mean density of the decorative particles to be no less than 10 per cm$^2$ for the reliable operation of PIVlab.

In our study, the polyamide particles on the liquid surface gather into about 0.1 mm-large tracer-complexes that have a rigid structure and are constantly moving under an external force. The tracer complexes form larger clusters, the structure of the latter maintains to be flat, but is strongly indented. The size of the clusters during the experiment gets larger, and at measurement time of over 10 hours their size can increase up to several centimeters, their quantity decreases by one order so that it is no more possible to use the PIV method \citet{Thielicke2014}. 

The clustering of particles is observed in many systems, for example, in the semiconductor bulk impurities segregation takes place [\cite{Kashammer2015}], also at grain boundaries [\cite{Straumal2016}]. At the same time, the motion of the dopants may be caused by an electric force or mechanical tension. The capillary interaction between floating particles on the liquid surface, which leads to their convergence, had been widely studied for many years, see for example [\cite{singh_joseph_2005}]. In our studies, polyamide-12 particles form homogeneously structured complexes under Van der Waals forces [\cite{Lifshitz1992}], while the clusters get formed by differently structured complexes. The complexes are maintained within the clusters by surface tension forces [\cite{Levchenko2019}]. The presence of background fluid flows on the surface provides a random convergence of the complexes.

\section{Experimental procedure}
The experimental procedure of generating vortex current by waves and visualizing liquid surface flows is described in detail in our paper [\cite{Filatov2018, Filatov2019}]. In our current experiments on the clusterization of polyamide-12 particles on the water surface, a 70-by-70~cm bath was used. The polyamide particles were sputtered above the water surface. Under the action of gravity, they descended onto the surface, forming various structures. Strong plunger pumping which can produce high-amplitude waves and vortices was used for creation of uniform distribution of rigid poliamide complexes over the surface and over the size. The velocity of the surface flow exceeded 10~cm/s. After the pumping was switched off, a series of photos of the water surface was taken at various intervals for a period of up to 120 hours. During the first 1.5 hours the shots were taken each 5 seconds, the rest of the time each next shot was made in 30 seconds. To remove the impact of the bath walls on measurement results, the number of the surface structures was calculated within the area of 50x50~cm$^2$ in size. The particles on the water surface were illuminated by a LED strip mounted on the side panes of the bath. The area of the structures was calculated from the number of pixels per surface structure.

Therefore, the experiments were carried out under different conditions, which gave rise to different average values of the background velocity, namely normal illumination (V $\sim$ 0.01~cm/s), switched off illumination (V $\sim$ 0.005 cm/s), a thermally isolated container (V $\sim$ 0.002 cm/s), and artificial operation of the plunger (0.006 to 0.25~cm/s).

\section{Experimental results}

\begin{figure}[ht] 
 \center
 \includegraphics [scale=0.5] {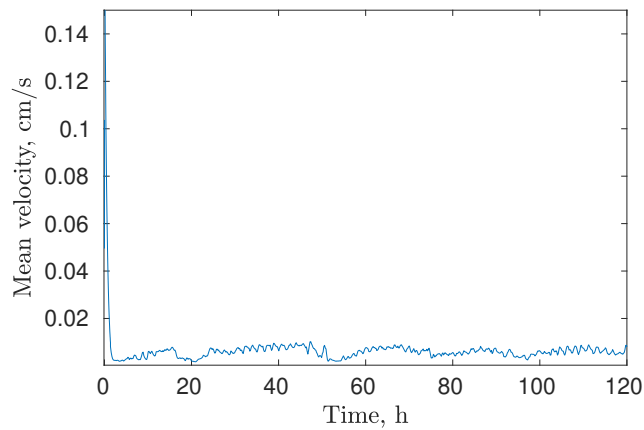}
 \caption{Time dependence of an average modulus of the flow velocity on the water surface.} 
 \label{img:meanVel} 
\end{figure}

As soon as the pumping is switched off, the surface flow velocity decreases due to the wave decay because of viscous loses. Fig. \ref{img:meanVel} shows the flow velocity reaching its minimum in several hours, slow random movements with the average velocity of the order of 0.01~cm/s remain on the surface afterwards. Such background movement on the surface maintains itself without plunger pumping, being caused by volume convectional flows that occur due to vertical and horizontal thermal gradients. The thermal gradients occur because of heat generation by the LED strip on the bath walls. 

\begin{figure}[ht]
 \begin{minipage}[ht]{0.24\linewidth}
 \center{\includegraphics[width=1\linewidth]{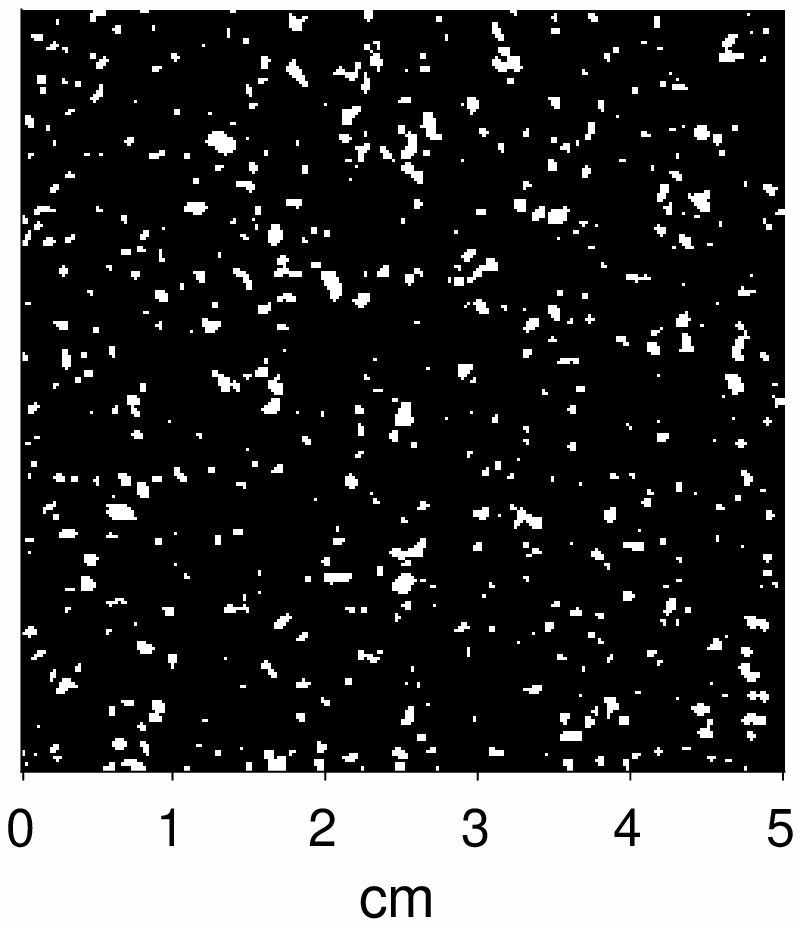} \\ a)}
 \end{minipage}
 \hfill
 \begin{minipage}[ht]{0.24\linewidth}
 \center{\includegraphics[width=1\linewidth]{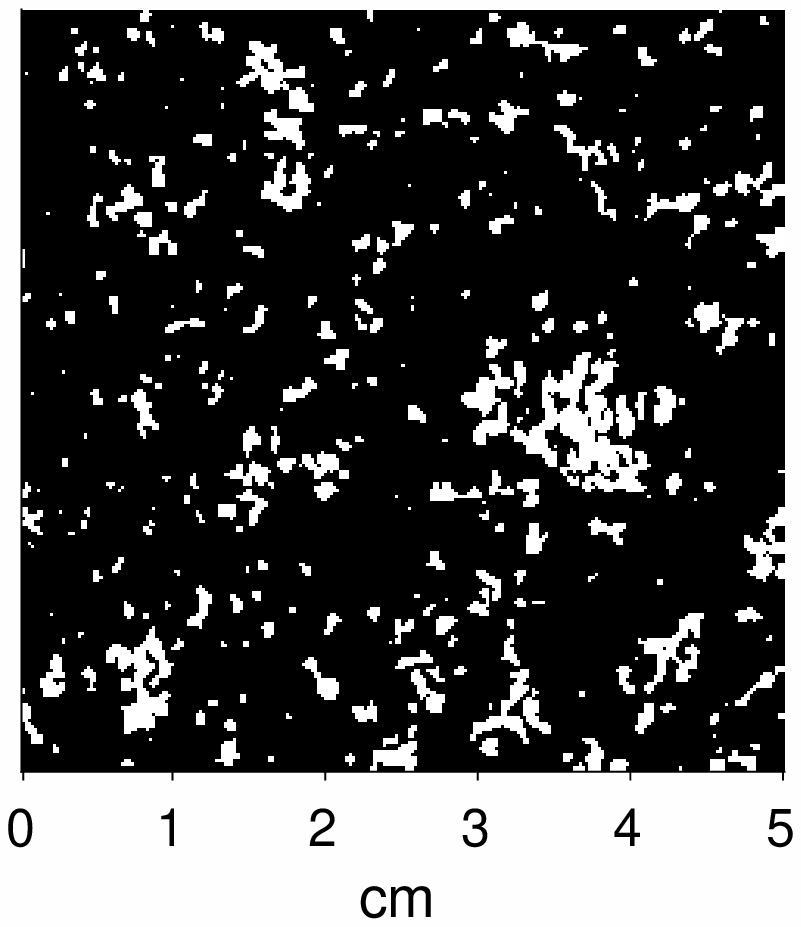} \\ b)}
 \end{minipage}
 \hfill
 \begin{minipage}[ht]{0.24\linewidth}
 \center{\includegraphics[width=1\linewidth]{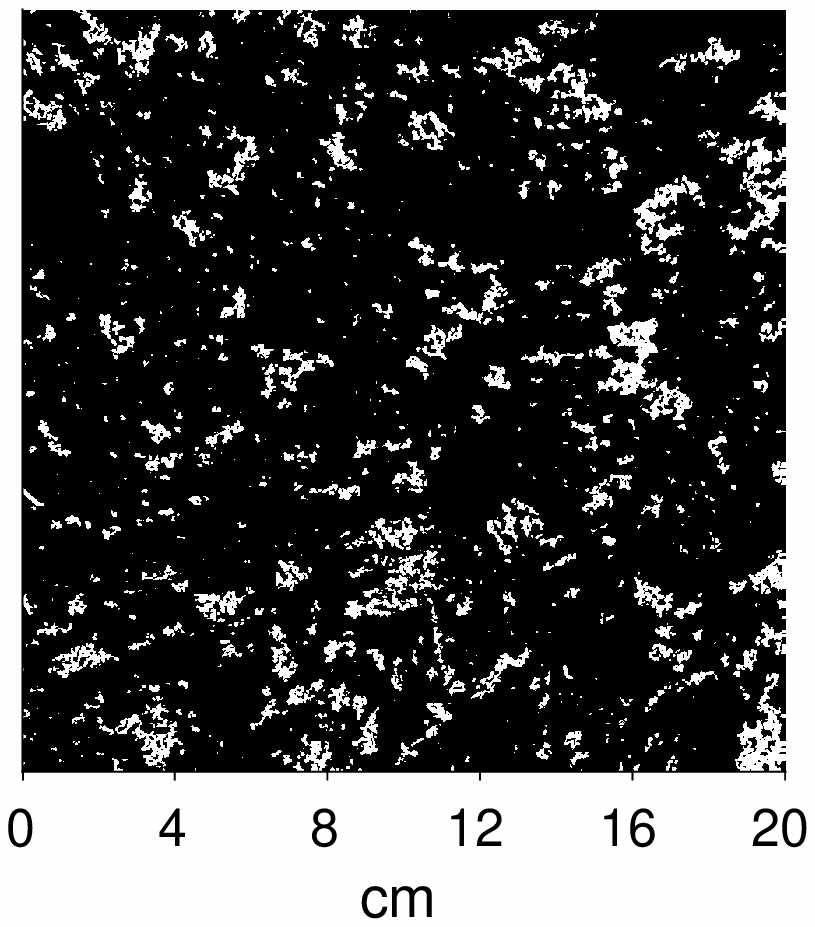} \\ c)}
 \end{minipage}
 \hfill
 \begin{minipage}[ht]{0.24\linewidth}
 \center{\includegraphics[width=1\linewidth]{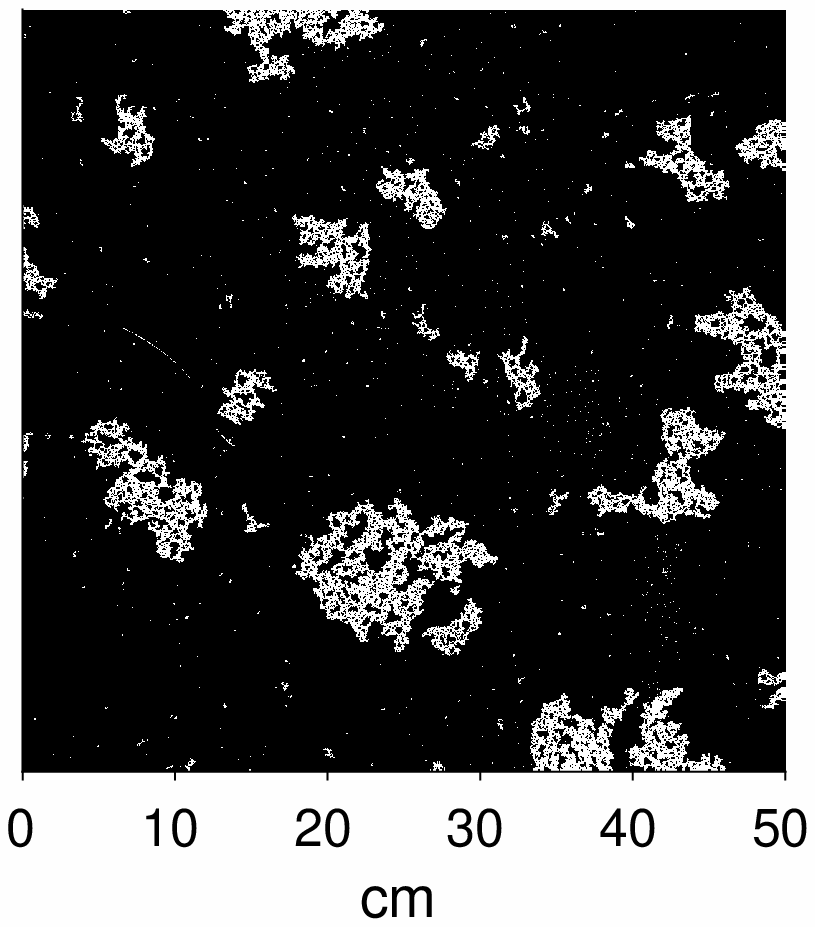} \\ d)}
 \end{minipage} 
 \caption{Photographs of structures on the water surface  10 minutes (a), 22 hours (b), and 60 hours (c) after switching off pumping. (d) is the photographs of the water surface 20 hours after switching off pumping without continuous illumination. }
 \label{img:photo} 
\end{figure}

Fig. \ref{img:photo} illustrates the photos of bath fragments with polyamide particle complexes on the water surface after 10 minutes (a), 22 hours (b), and 60 hours (c) after switching-off pumping. Initially, after switching off pumping, there are approximately 10$^5$ complexes with an area of the order of 10$^{-4}$ - 10$^0$~cm$^2$ on the surface, formed by polyamide microparticles.

As it can be seen, most complexes are disc-shaped (\ref{img:photo}a). However, we can observe the processes of the merging of two disc-complexes into one, as well as of three discs into linear I- and T-shaped complexes. Over time, the complexes formed at the first consolidation stage are collected into larger formations with a linear size of up to 0.3~cm, and this process continues for tens of hours, Fig. \ref{img:photo}b. At this stage the complexes are characterized by their surface structure homogeneity. In 60 hours, the complexes form clusters which have a complex branched structure of several centimeters in size.

This can be seen more clearly in Fig. \ref{img:photo} (d) that shows complexes on the water surface 20 hours after switching off the pumping and without continuous illumination by the LED strip. The background convective flow velocity got halved, reducing to 0.005~cm/s. The photo demonstrates clearly that the clusters size is larger than that in Fig. \ref{img:photo}(c) that was taken after 60 hours of clusterization with the heat source activated.
We can conclude that the clustering process on the water surface has two stages. At the first stage, the continuous complexes grow up to their characteristic size of less than 0.5~cm. At the second stage the continuous complexes form clusters with a complex branched structure. The consolidation rate of the complexes and their maximum size are determined by the surface convective flow.

\begin{figure}[ht]
 \begin{minipage}[ht]{0.32\linewidth}
 \center{\includegraphics[width=1\linewidth]{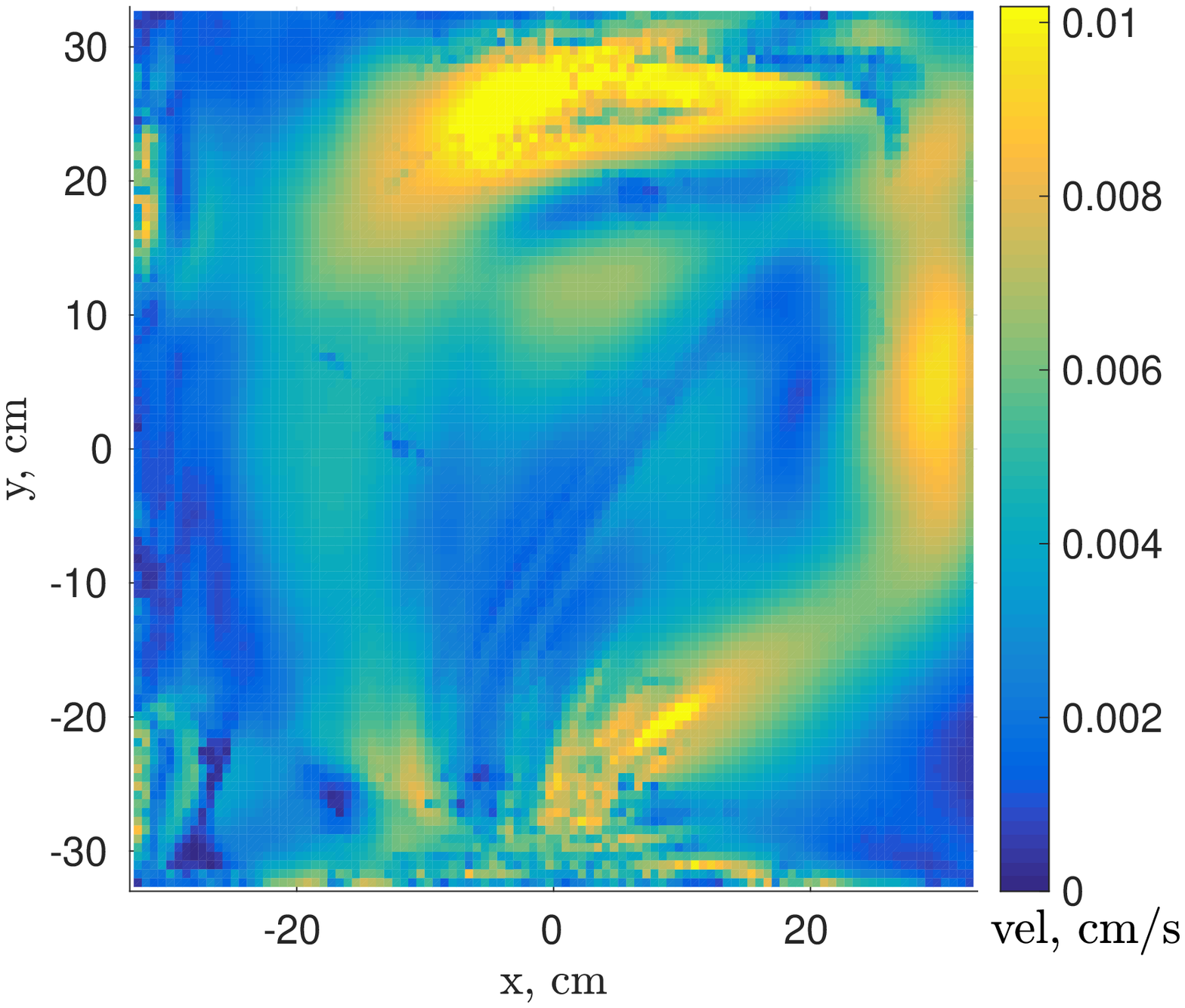} \\ a)}
 \end{minipage}
 \hfill
 \begin{minipage}[ht]{0.32\linewidth}
 \center{\includegraphics[width=1\linewidth]{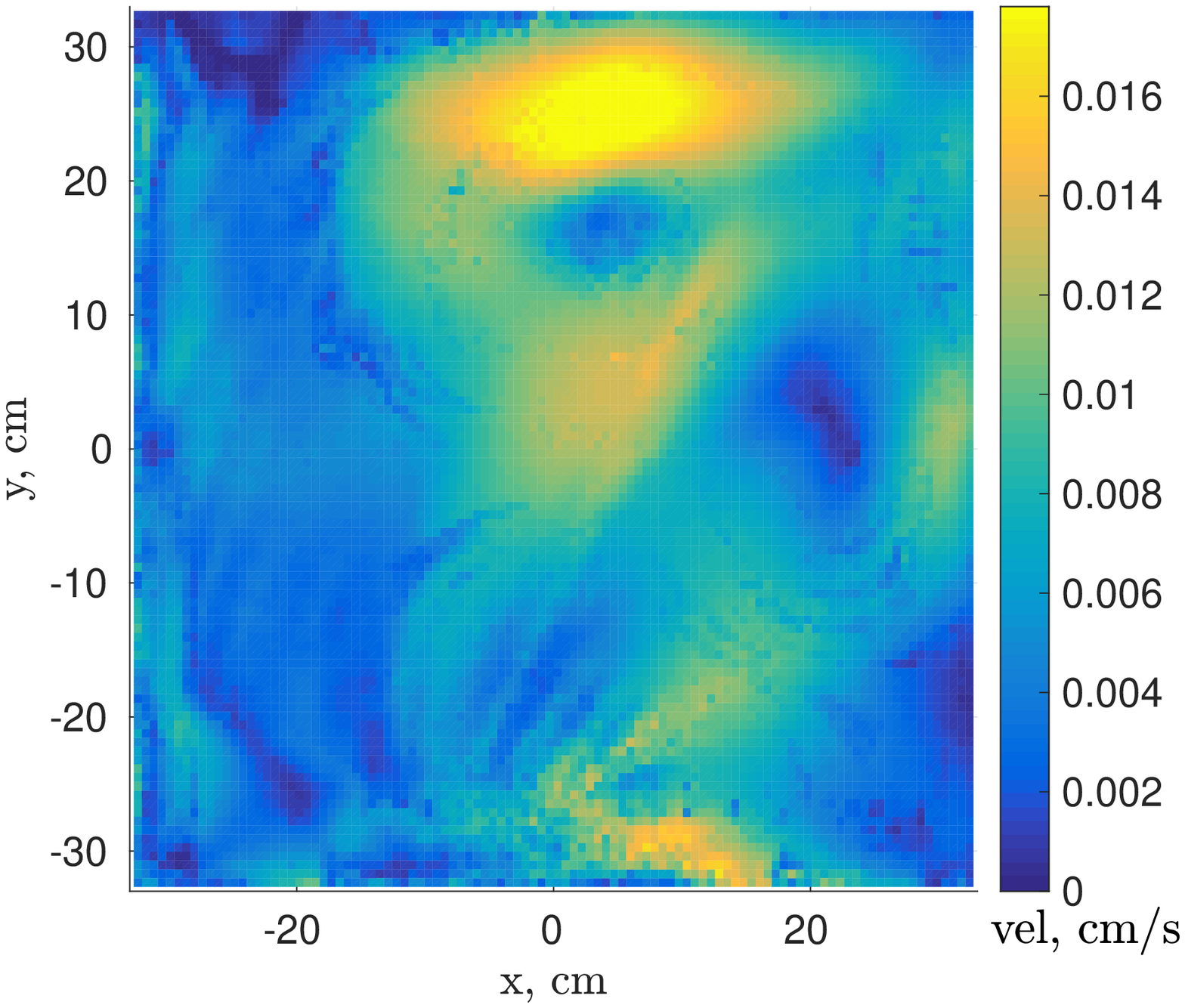} \\ b)}
 \end{minipage}
 \hfill
 \begin{minipage}[ht]{0.32\linewidth}
 \center{\includegraphics[width=1\linewidth]{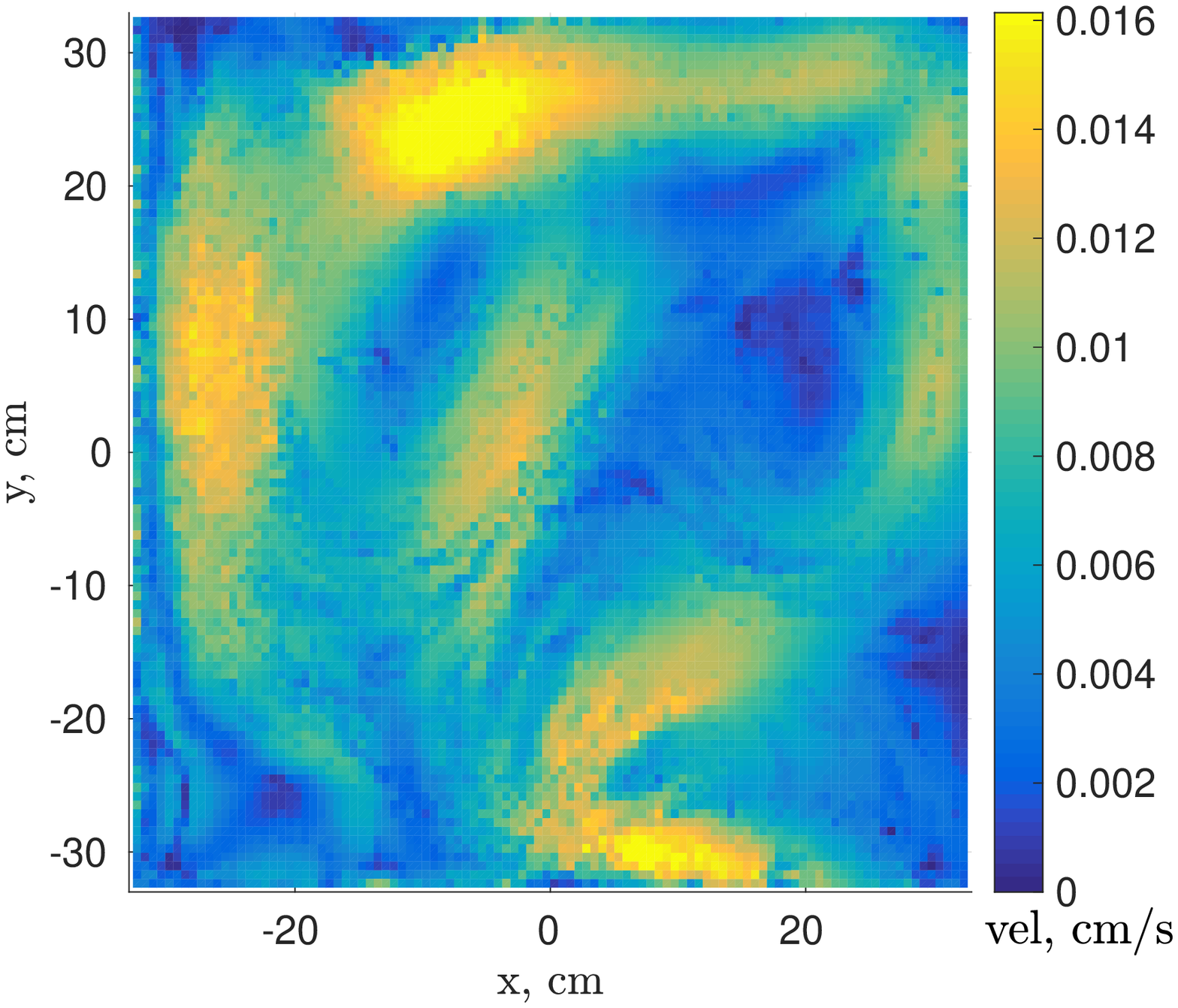} \\ c)}
 \end{minipage}
 \vfill
 \begin{minipage}[ht]{0.32\linewidth}
 \center{\includegraphics[width=1\linewidth]{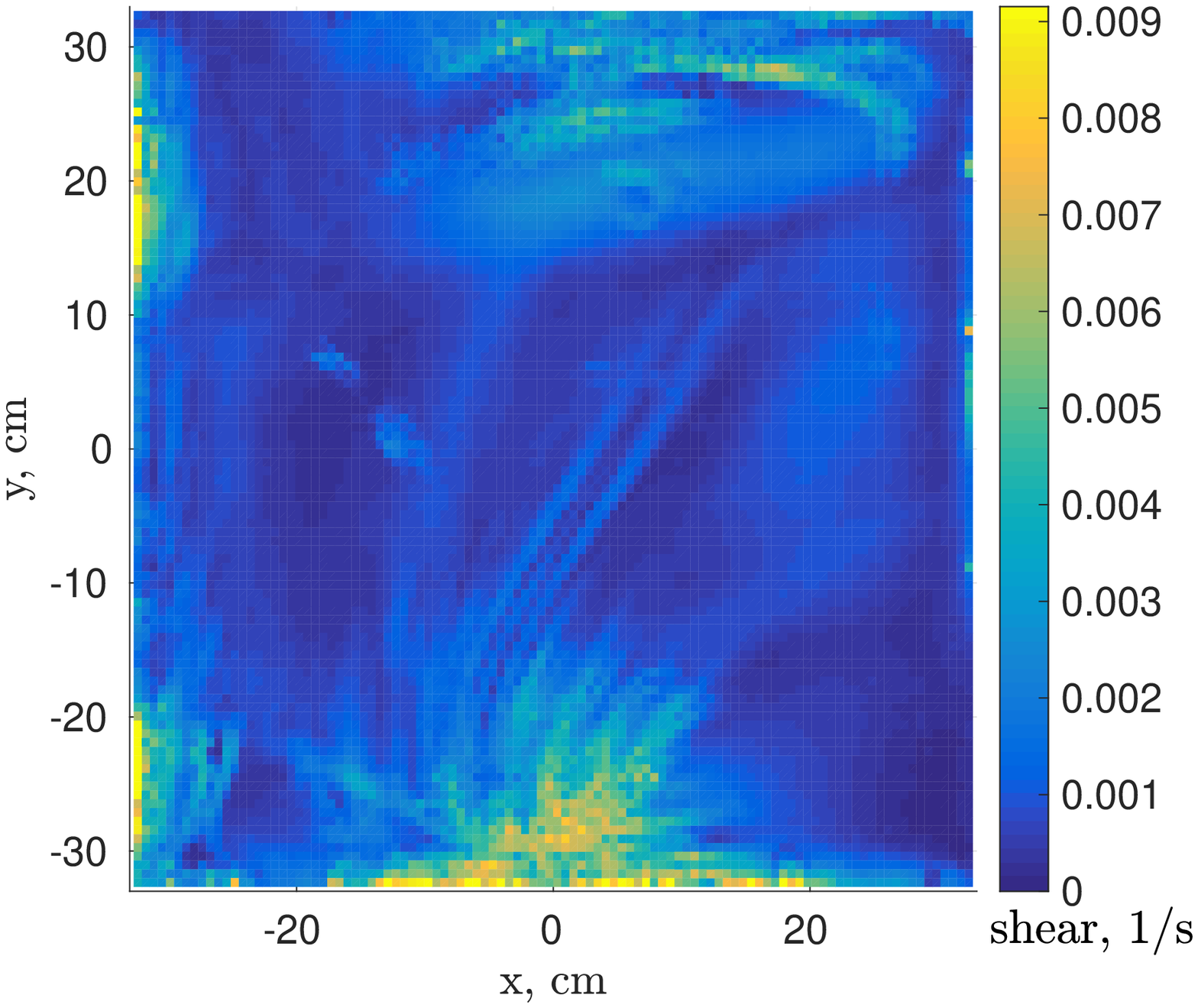} \\ d)}
 \end{minipage}
 \hfill
 \begin{minipage}[ht]{0.32\linewidth}
 \center{\includegraphics[width=1\linewidth]{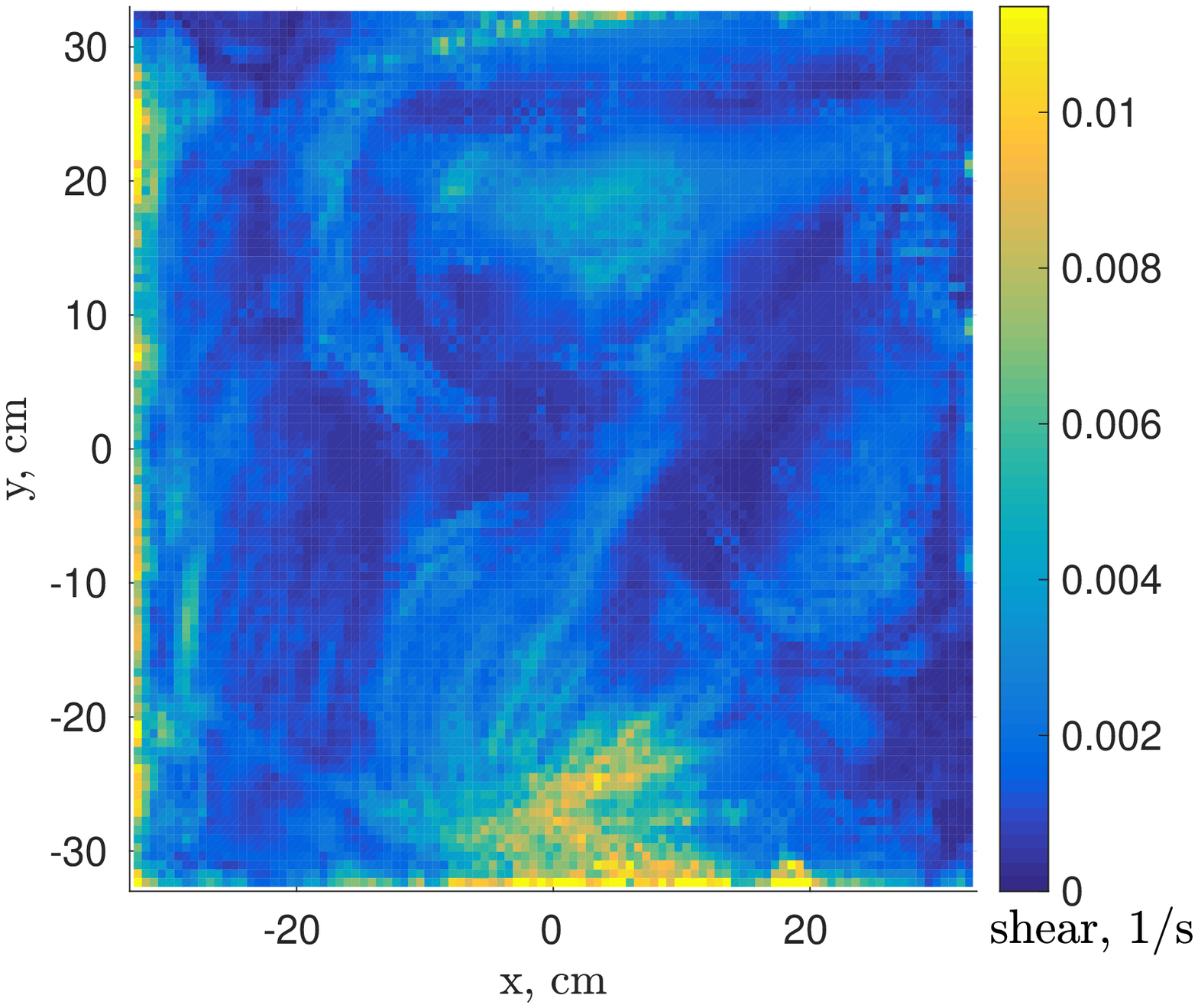} \\ e)}
 \end{minipage}
 \hfill
 \begin{minipage}[ht]{0.32\linewidth}
 \center{\includegraphics[width=1\linewidth]{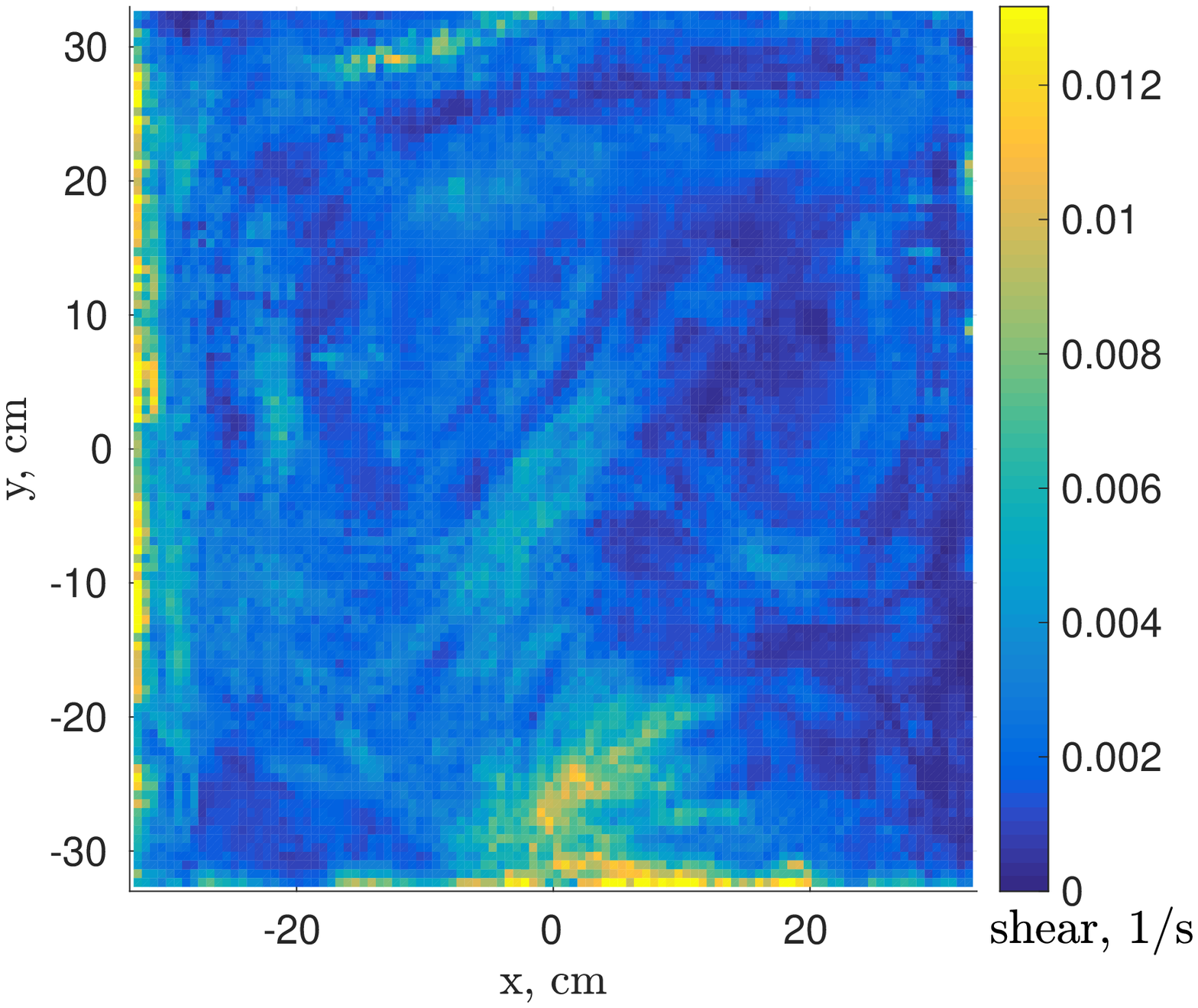} \\ f)}
 \end{minipage}

 \caption{Distribution of absolute velocity over the surface 20 hours (a), 60 hours (b) and 86 hours (c) after switching off pumping. Distribution of particular velocity gradients (shear) $\sqrt{(\partial V_x / \partial y)^2 + (\partial V_y / \partial x)^2} $ over the surface 20 hours (d), 60 hours (e), and 86 hours (f) after switching off pumping.}
 \label{img:velgrad} 
\end{figure}

Fig. \ref{img:velgrad} illustrates the distribution of absolute velocity and velocity gradients (shear) over the surface 20, 60 and 86 hours after switching off pumping. One can see in the images that velocity distribution over the surface changes significantly with time. This is because heating conditions leading to the emergence of convection are non-stationary and non-uniform. It may also be seen from the figure \ref{img:velgrad} (a-c) that the characteristic correlation lengths for the velocity are much larger than the size of the clusters.

\begin{figure}[ht] 
 \center
 \includegraphics [scale=0.5] {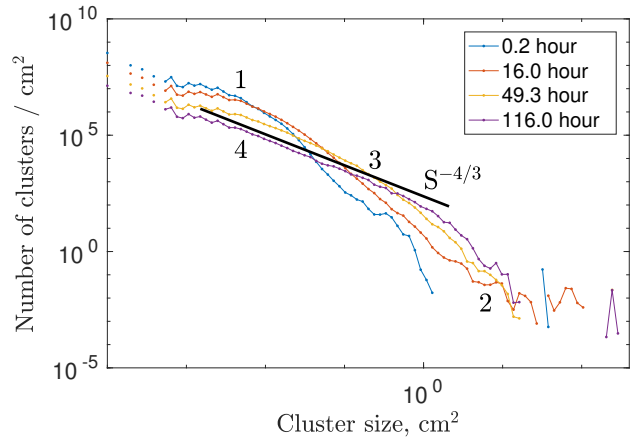}
 \caption{ Distribution of the normalized number of structures on the water surface over their area $N(S)$ after 12 minutes (1), 16 hours (2), 49.3 hours (3), and 116 hours (4) after switching off pumping. The average surface flow velocity is 0.01~cm/s. } 
 \label{img:hist128h} 
\end{figure}

Fig. \ref{img:hist128h} shows the distributions of the normalized number of structures on the surface over their area $N(S)$, obtained by dividing the number of complexes and clusters from the photos, the area of which fits within the specified interval, by the number of photos and interval width. The interval increases with a logarithmically increasing step from 10$^{-4}$~cm$^2$ to 50~cm$^2$, the number of photos is 10$^3$ units. Curve 1 was obtained 12 minutes, Curve 2 16 hours, Curve 3 49.3 hours, and Curve 4 116 hours after switching off pumping. It is seen that the probability of finding large clusters on the photos is less than unit.

Immediately after switching off pumping (see Curve 1), when the surface flow velocity is still high, most surface structures have an area of about 10$^{-2}$~cm$^2$ or less. Only a few of them have an area of about 1~cm$^2$. At 16 hours after switching off pumping (see Curve 2), when the wave motion has totally decayed, the photo shows many structures in wide range of areas of up to 2~cm$^2$. Nevertheless, in distribution 2 has not reached its equilibrium state, therefore, it cannot be described by an universal function. At 49.3 hours (see Curve 3), the large cluster growth continues by joining smaller complexes. Clusters of about 10~cm$^2$ in size appear. At 116 hours after switching off pumping we observe that the distribution of the surface structures in a broad range of areas from 0.001~cm$^2$ up to 1~cm$^2$ is getting close to the power-law function. It is clearly seen that the obtained distribution $N(S)$ can be described by the $N(S) \sim (S/S_0)^{-n}$ function with an exponent close to n = 4/3. However, we should notice that the $N(S)$ distribution at $S > 1$~cm$^2$ significantly deviates from the power-law function. The reason for this may be too high velocity of the background surface water flow, or lack of the observation time.

Thus, the distribution of the normalized number of surface structures over their area $N(S)$ at long observation time becomes a power-law one, with its characteristic size $S_0 \approx 2 \cdot 10^{-2}$~cm$^2$.

\begin{figure}[ht] 
 \center
 \includegraphics [scale=0.5] {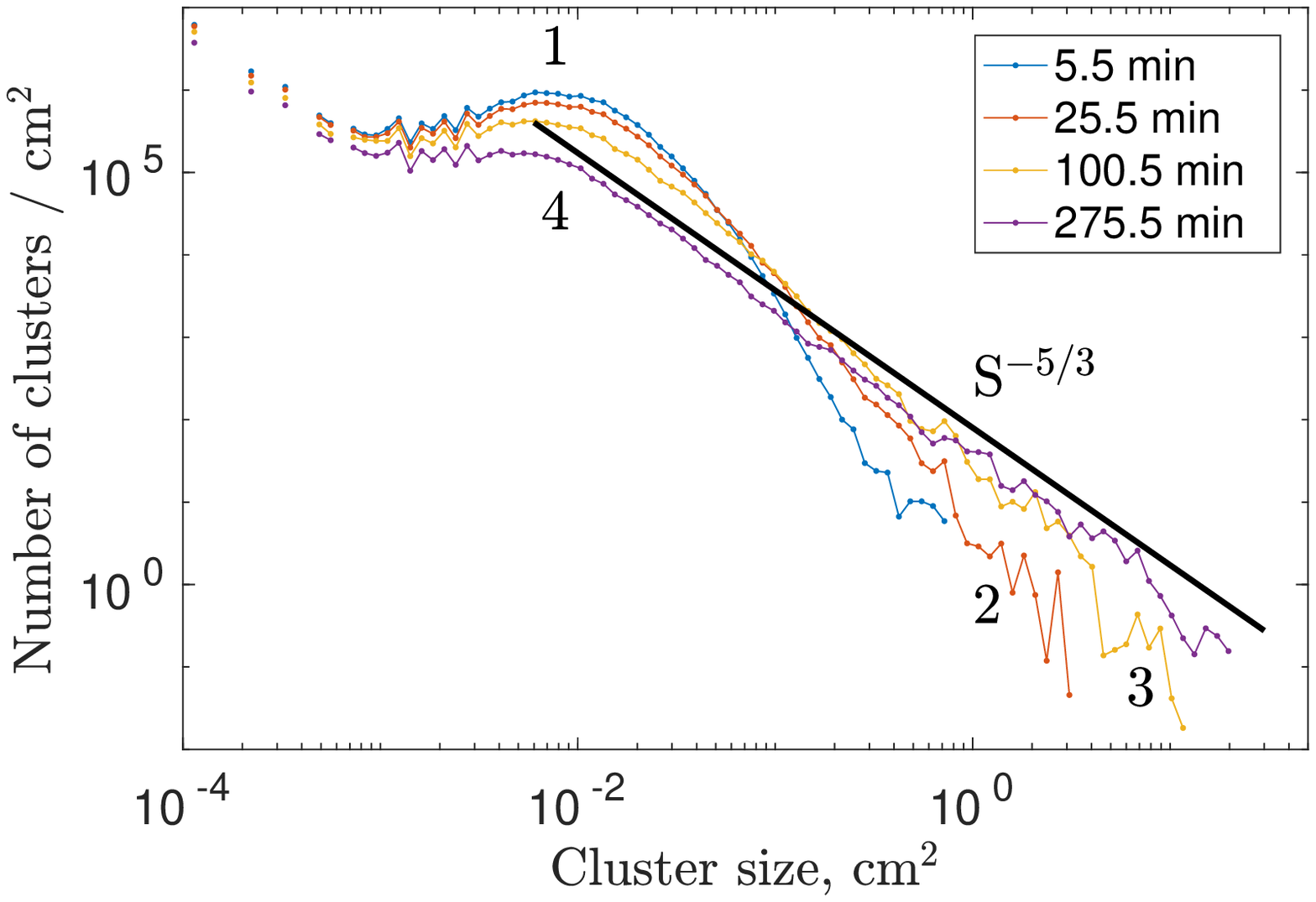}
 \caption{ Distribution of the normalized particle cluster number the size $N(S)$ in 5.5 minutes (1), 25 minutes (2), 100.5 minutes (3), and 275.5 minutes after switching off pumping. The average surface flow velocity is 0.002~cm/s.} 
 \label{img:hist500m} 
\end{figure}

As we mentioned above, the average value V of random background flow velocity on the surface can define the characteristic time of the $N(S)$ function reaching its equilibrium state, as well as the maximum size of clusters. To test this assumption, the average velocity V was reduced to 0.002~cm/s by setting additional 5-cm thick Styrofoam insulation on the bath walls and changing the illumination mode. Fig. \ref{img:hist500m} shows the distributions $N(S)$ obtained 5.5 minutes, 25 minutes, 100.5 minutes, and 275.5 minutes after switching off pumping at frequency of 3 Hz. The mean surface current velocity immediately the pump at a frequency of 3 Hz was switched off getting close to 10 cm/s

It is clearly seen that the process of the establishment of a power-law distribution proceeds much quicker than in the case of high values of background flow velocity. The power-law distribution is already observed in about 1.5 hours (see Curve 3).

In an area range from 10$^{-1}$~cm$^2$ to 10$^1$~cm$^2$, the observed dependence $N(S)$ is described by a power function with an exponent close to n = 9/4. It can be seen that the number of clusters increased due to a decrease in the number of smaller clusters. Curve 4 obtained in 4.6 hours in an area range from 10$^{-2}$ to 10$^1$~cm$^2$, can be well described by the power function $N(S) \sim (S/S_0)^{-n}$ with an exponent close to n= 5/3 and $S_0 = 1.0 \cdot 10^{-1}$~cm$^2$. The process of redistribution of areas between small and large structures leads to a decrease in the absolute value of the exponent of the function describing the experimental dependences $N(S)$. The probability of discovering a large cluster on the water surface increases with time.

\begin{figure}[ht]
 \begin{minipage}[ht]{0.49\linewidth}
 \center{\includegraphics[width=1\linewidth]{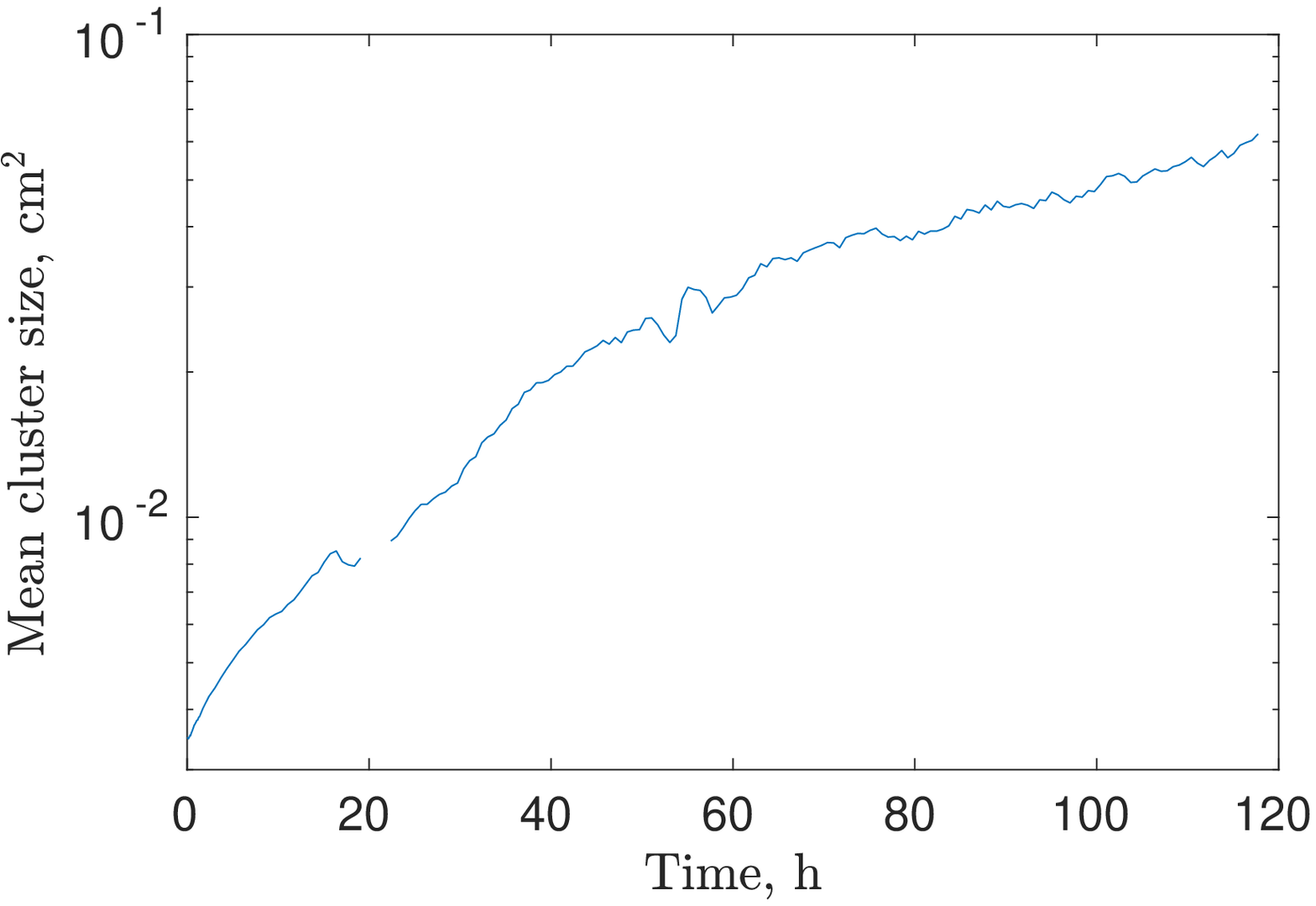} \\ a)}
 \end{minipage}
 \hfill
 \begin{minipage}[ht]{0.49\linewidth}
 \center{\includegraphics[width=1\linewidth]{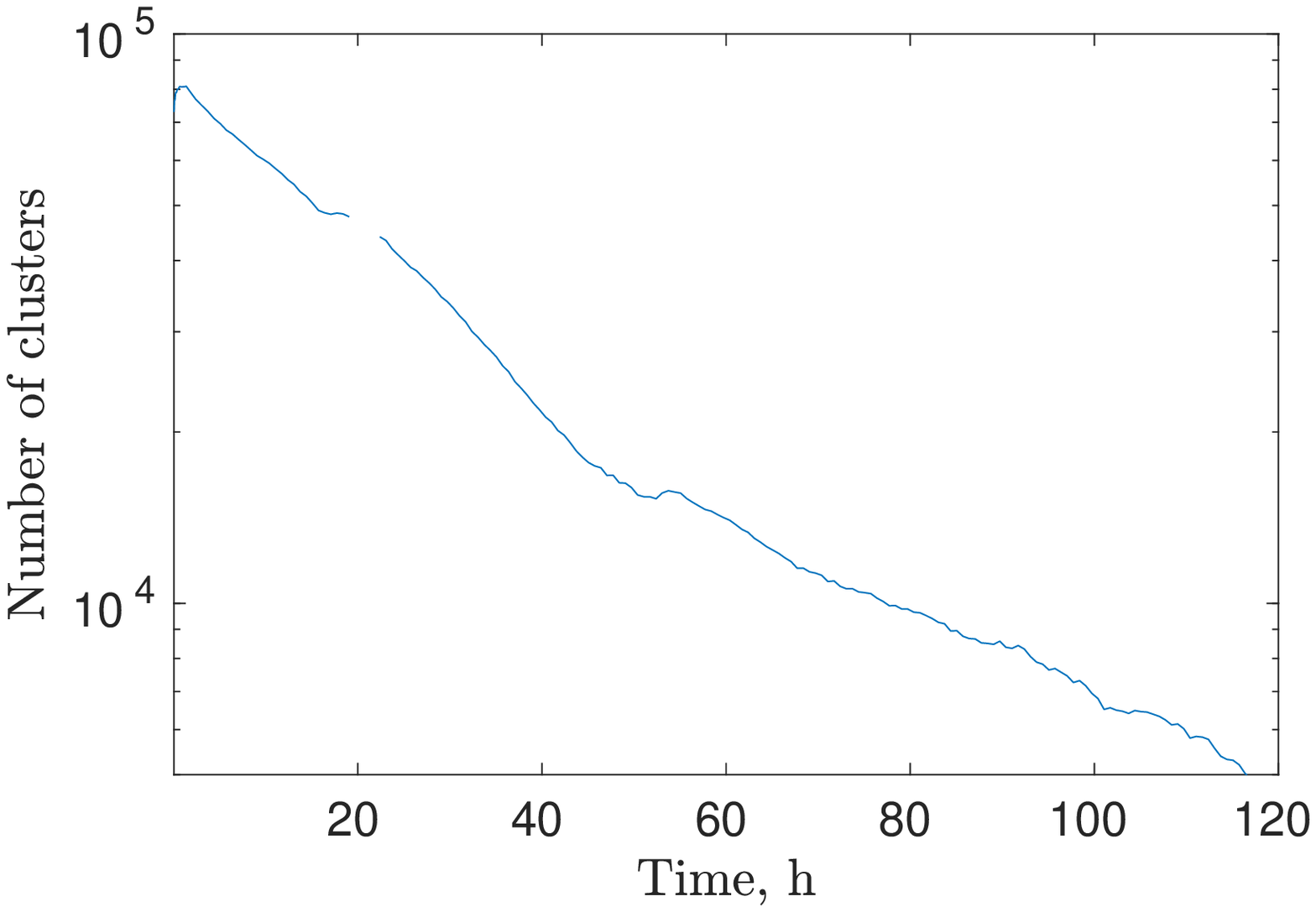} \\ b)}
 \end{minipage}
 \caption{Time dependence of an average cluster size (a) and the number of clusters (b) on the water surface 50x50~cm$^2$ on time after switching off pumping. The background liquid flow is 0.01~cm/s.}
 \label{img:clusterSize} 
\end{figure}

The maxima on the distributions $N(S)$ (see Curves 1, 2 and 3) are clearly seen. One can assume that the concentration of complexes near $S \approx 10^{-2}$~cm$^2$ within the first minutes of observation is caused by intensive plunger pumping that makes large clusters break into smaller ones by intensive fluid flows. 
Fig. \ref{img:clusterSize}(a) shows the time dependence of an average area of structures in semi-logarithmic coordinates after switching off excitation at background flow velocity of 0.01~cm/s. The average size of structures, defined as the sum of the areas of all complexes and clusters $S$, divided by their number $N$, increases with time, $S_{EV} = S/N$. One can see that, although 116 hours had elapsed since switching off excitation, the clustering process was not completed. We assume that during the first 40 hours the average area $S_{EV}$ of complexes and clusters increases exponentially within a characteristic time of $ \tau_0 \sim 15 $ hours. Starting from the 45th hour, an average size of a cluster increases exponentially within a characteristic time of $ \tau_0 \sim 40 $ hours as a result of the linear structure fusion. 

Fig. \ref{img:clusterSize} (b) demonstrates the time dependence of the number of clusters, in semi-logarithmic coordinates as well, after switching off pumping. It is obvious that the number of clusters $N(t)$ decreases from $8 \cdot 10^4$ units to $5 \cdot 10^3$ with time. At times of less than 45 hours, the decreasing process is approximately exponential, its characteristic time being $ \tau_0 \approx 15$ hours. At times of over 45 hours, the process of a decrease in the cluster number is also exponential, $N(t) \sim exp(-t/\tau_0), \tau_0 \approx 40$ hours. The difference between the characteristic times is due to the qualitative change in the structures that form complexes and clusters. In the beginning, the formation of linear structure-complexes of large area, involving the disc-complexes, seems to be the main process. But at long times, the growth of clusters is associated mainly with the merging of linear surface structures.

The comparison of Fig. \ref{img:clusterSize} (a) and \ref{img:clusterSize} (b) shows clearly that an average size of clusters increases due to a decrease in the total number of structures while their total area $S$ remains the same. 

\section{Discussion}

The power-law distribution of probability of a $C/x^n$-type event is called the Pareto distribution [\cite{Manceau2019}] and is observed in economics [\cite{Axtell2001}], geophysics [\cite{Taubert2018}], astrophysics [\cite{Edward1991}], and condensed-matter physics [\cite{Stefani2009, Manceau2019}]. It should be noted that the Pareto distribution is often formed in random interactions in system of limited resources, such as the income of money per citizen of a country or distribution of cities over the number of inhabitants [\cite{Jones2015, RibeiroJusto2014}]. In our experiments, the total surface structure area is limited by the total number of polyamide-12 particles sputtered on the water surface, and it cannot exceed the area of our experimental bath. The randomness of interaction is defined by a background flow on the surface water. 

In our studies, the polyamide-12 particles had been already merged into the complexes due to the action of Van der Waals forces. We assume that an average density of the complexes turned out to be lower than that of water due to hollow spaces that occur in their bulk. So, therefore, after mixing by intensive pumping, they are immersed in water, and a part of them with the height $h$, completely moistened with a liquid, protrudes above the surface. By interacting with each other under the liquid surface by van der Waals forces, the complexes are assembled into flat discs with a characteristic size of about 0.2-0.3~cm and a thickness of about 0.01~cm. Such complexes include up to 10$^6$ polyamide-12 particles. This is evidenced by the predominant number of complexes in Figs. \ref{img:hist128h} and \ref{img:hist500m} with areas of less than 10$^{-2}$~cm$^2$.

The disc-complexes with a continuous surface structure seemingly cannot reach larger sizes due to the rigidity of the structure, the directional action of Van der Vaals forces, and the destructive effect of the flows spreading on the surface. The total number of smaller complexes significantly exceeds the number of larger ones, so the process of increase in the average area of the surface structures is mainly related to the merging of the smaller complexes - the first stage. One can assume that the process of structure merging has an accidental character, so the velocity of changes in the number of clusters is proportional to their quantity $N$:

\begin{equation}
 \label{eq:velN}
dN(t)/dt = 1/\tau_0 \cdot N
\end{equation}

At the second stage, decrease in the number of clusters is mainly due to the addition of the smaller linear complexes to the larger clusters. Within very long consolidation times, when the number of clusters drops to just few, the exponential dependence will no longer make sence. 

Let us take a qualitative look at the process of merging of two Van der Vaals disc-complexes that have formed into one complex, that is, a linear structure, at the first clusterization stage. The converging of the two discs to the distance of the same order of their radius is performed by the background surface liquid flow. The values of the disc diameters are close to the capillary length $\sqrt{2\sigma / (\rho g)}$ , which is equal to 3.9 mm [\cite{Landau1987}] for water, $\sigma$ is the surface tension coefficient, $g$ is the free fall acceleration, and $\rho$ is the water density. Therefore, we assume that the process of complex merging at this stage is defined by the action of surface tension forces. 

Let us assume that in the formed complex the distance between the centers of the disks is equal to the sum of their radii, $R_1 + R_2$. We also assume that the thickness and the total complexes area do not change when they get merged. So the energy balance between only the energy change in the side surface of the complexes and the height $h$ should be taken into account. The sum of side surface energies of two different discs can be written as:

$E_S = 2\pi h \sigma (R_1 + R_2)$

After merging into a dumbbell shape, the energy of the side surface will be written as:

$E_L \approx h \sigma (\pi + 2) (R_1 + R_2)$

It can be seen that the energy ratio $E_S/E_L \approx 2 \pi /(\pi + 2)$ always exceeds unit. To prove our suggestions, let us look at Fig. \ref{img:photo} (b) that shows a large quantity of linear complexes which is the evidence of the initial prevailing of the disc complexes that do not change after merging. The formation of large disc-shaped complexes at this stage is rare.

At the next stage, large clusters with a complicated structure get formed. Their "building material" is complexes of various shapes: discs, dumbbells, I- and T-shaped structures. As a result of the merging, complicated structures get formed that do not have any symmetry, with broken boundaries and holes in their surface (See Fig.  \ref{img:photo}(c) and (d)). The clusters drift on the water surface and retain their shape for a long time. Changes occur when they merge with a smaller complex, or two larger clusters merge together.

We should note that the process of merging with a complex and merging of clusters are possible only when there is a chaotic liquid flow with the average velocity $V$. However, the presence of very high velocities of the background fluid flow will prevent an increase in the cluster sizes.

It follows from simple considerations that the joining of a small complex of the radius $R_S$ to a large cluster of the size $R_B$ is possible if a decrease in the total energy of the large plus small clusters exceeds the kinetic energy of the small cluster.

The energy change in the united larger cluster is approximately equal to:

$\Delta E \approx (\pi -2) h \sigma R_S$

The cluster kinetic energy equals to $M_SV^2/2$. The cluster mass equals to $M_S = \pi R_S^2 h \rho_c$, where $\rho_c$ is the cluster density. 

Thus, the condition for joining a smaller cluster to a larger one is expressed in the inequation:

$(\pi-2) h \sigma R_S > \pi R_S^2 h \rho V^2$,

from which we get qualitative size restriction for the smaller complex that can merge with a larger cluster at a given liquid flow velocity V.

$R_S < (\pi-2)/\pi \sigma/ \rho V^2$

Thus, the growth of larger clusters is due to the joining of mainly smaller complexes to them. The burnout of smaller complexes of less than 0.3~cm in size occurs mainly on the surface (see Figs. \ref{img:hist128h} and \ref{img:hist500m}). 

However, the size $R_B$ of a larger cluster is also restricted by the action of forces caused by shear liquid flows that break the bonds holding the complexes together. Let us assume that the velocity gradient $\Delta V$ is order to $V/R_B$ in order of magnitude, the cluster area is $ \sim R_B^2$, then for the estimation we can write:

\begin{equation}
 \label{eq:largeClusterSize}
V/R_B \cdot R_B^2 \cdot \eta \sim \sigma R_S
\end{equation}

From (\ref{eq:largeClusterSize}) we find that the maximum cluster area with an increase in the velocity of the background shear flow should decrease according to the law:

\begin{equation}
 \label{eq:largeClusterSize1}
 R_B^2 \sim (\sigma R_S / V / \eta)^2,
\end{equation}

where $\eta$ is the coefficient of dynamic viscosity.

\begin{figure}[ht] 
 \center
 \includegraphics [scale=0.5] {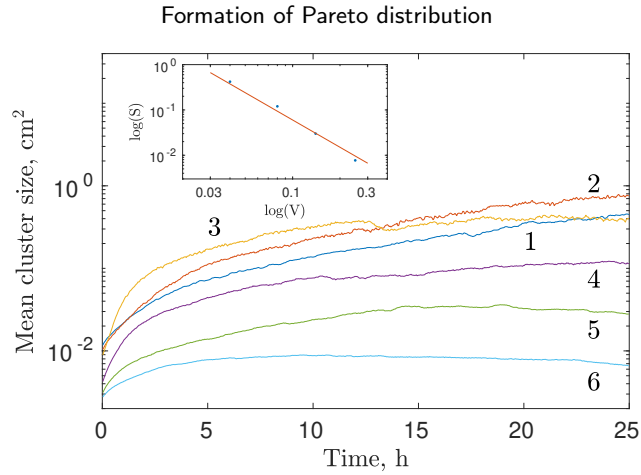}
 \caption{Time dependence of average cluster area on the water surface $S_{EV}$ on time at various background flow velocity values in semi-logarithmic coordinates. The insert demonstrates dependence of the maximum average cluster area on the surface flow velocity in a log-log scale. 1 - 0.006~cm/s, 2 - 0.01~cm/s, 3 - 0.04~cm/s, 4 - 0.08~cm/s, 5 - 0.14~cm/s, 6 - 0.25~cm/s.
} 
 \label{img:SEV} 
\end{figure}

To testify this dependence, we investigated the consolidation process at various plunger operating regimes. The mean background flow velocity $V$ of the liquid changes almost by two orders of magnitude from 0.006 to 0.25~cm/s. Fig. 6 shows the dependencies of an average area of the structures at various background velocities. It can be seen that at velocities of 0.01 and 0.006~cm/s an average cluster size continually and exponentially increases with time throughout 25 hours of investigation. At a velocity of 0.04~cm/s and higher, the value of $S_{EV}$ reaches its maximum value, moreover, the higher the velocity $V$, the less the maximum value of $S_{EV}$. It is surprising that the experimentally measured average cluster size decreases in an inverse ratio to the squared velocity of the background flow according to formula (3) (See the insert in Fig.\ref{img:SEV}).

Note that it was shown earlier in [\cite{Higashitani1998}] that for clusters with a fractal structure, the average size of a stable aggregate decreases with an increase in the shear flow velocity by a power law with the exponent less than unity in modulus. However, for clusters with closer packing [\cite{Vassileva2006}], the average size of an aggregate slightly depends on the shear flow velocity. The long-term dynamics of the shear-induced breakage of individual colloidal clusters covering a wide range of fractal dimensions were studied in [\cite{Harshe2016}].

\begin{figure}[ht] 
 \center
 \includegraphics [scale=0.5] {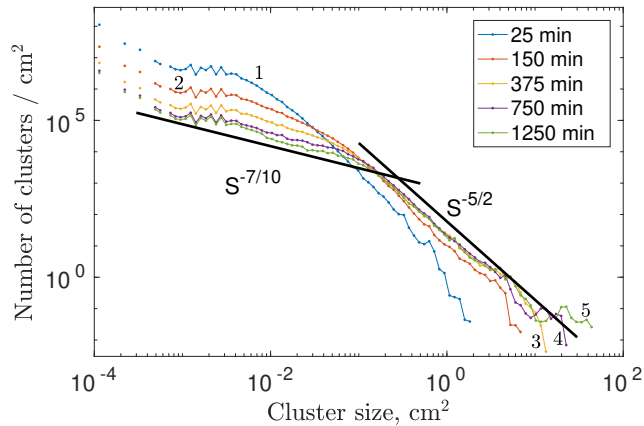}
 \caption{Distribution of the normalized number of clusters over the surface area $N(S)$ 25.2 minutes (1), 150 minutes (2), 375 minutes (3), 750 minutes (4), and 1250 (5) minutes after switching off excitation.} 
 \label{img:hist1200m} 
\end{figure}

Fig.6 shows that the liquid flow with a velocity of 0.04~cm/s and higher significantly restricts cluster growth. Note that at all values of the background liquid flow velocity, power-law distribution of the normalized number of clusters over the surface is observed. As an example, Fig. \ref{img:hist1200m} illustrates the results obtained at a flow velocity of 0.08~cm/s.

At 25 minutes switching off pumping, the distribution $N(S)$ is far from being a power-law type. However, starting from the 150th minute, on the distributions $N(S)$ we can clearly distinguish two regions that are described by the power functions $ \sim S^{-n}$ with the power exponent $n$ being approximately 7/10 and 5/2. The presence of two power-law distributions indicates different conditions for the formation of surface structures and proves our assumption that there are two consolidation stages, i.e., the formation of large complexes and formation of branched clusters from them. 

\section{Conclusion}

It has been experimentally determined that in the system of structures formed on the water surface by polyamide particles, the Pareto distribution is established in a wide range of their areas. Long-term studies have been carried out (up to 120 hours) of the process of transformation of macroscopic quasi-two-dimensional tracers that get formed from polyamide tracers used for visualizing the water surface flows in a pool of a limited size. The tracers consist of light neutral polyamide particles with a mean diameter of $\sim$ 30 $\mu km$. The finely dispersed polyamide-12 powder is sputtered onto the water surface in a bath, so that the uncovered water area exceeds the area covered with the dopant clusters by more than 100 times. During the experiment, the total area of the dopant clusters on the water surface remained unchanged. It observed that at long-term observation, the distribution of the normalized density of quasi-two-dimensional polyamide clusters can be described by the power function of $N(S/S_0) \sim C(S/S_0)^{-n} $ - Pareto distribution, where $S$ is the area of a single cluster surface, $S_0$ is the characteristic cluster area. The formation of two-dimensional structures on the water surface in a bath depends largely on the action of surface tension forces. The number of surface structures decreases exponentially while their total area remains constant. It has been also shown experimentally that the background liquid flow on the surface plays the key role in the clusterization process.

\section{Acknolgement}
We are grateful to V. Lebedev, I. Kolokolov, and V. Parfenyev for valuable discussions.
This work was supported by the Russian Ministry of Science and Higher Education, project No. 075-15-2019-1893.
\bibliographystyle{cas-model2-names}
\bibliography{cas-refs}

\end{document}